\documentclass[dvipdfmx]{aa}
\usepackage{natbib}
\bibpunct{(}{)}{;}{a}{}{,}

\usepackage{graphicx}
\usepackage[varg]{txfonts}
\def\diff{{\rm d}}

\begin{document}

   \title{Surface-effect corrections for solar-like oscillations using 3D hydrodynamical simulations}
   \titlerunning{Surface-effect corrections for solar-like oscillations using 3D hydrodynamical simulations}
   \subtitle{I. Adiabatic oscillations}

   \author{T. Sonoi\inst{1}
          \and
          R. Samadi\inst{1}
          \and
          K. Belkacem\inst{1}
          \and
          H.-G. Ludwig\inst{2,3}
          \and
          E. Caffau\inst{3}
          \and
          B. Mosser\inst{1}
          }
   %\runningauthor{}

   \institute{LESIA, Observatoire de Paris, PSL Research University, CNRS, Universit\'e Pierre et Marie Curie,
 Universit\'e Denis Diderot,  92195 Meudon, France
     \and
     Zentrum f\"ur Astronomie der Universit\"at Heidelberg, Landessternwarte, K\"onigstuhl 12, D-69117 Heidelberg, Germany
     \and
     GEPI, Observatoire de Paris, PSL Research University, CNRS, Universit\'e Denis Diderot, Sorbonne Paris Cit\'e, 5 Place Jules Janssen, 92195 Meudon, France
             }

   %\offprints{T. Sonoi}
   %\mail{takafumi.sonoi@obspm.fr}
   \offprints{\tt takafumi.sonoi@obspm.fr}
   \date{\today}

% \abstract{}{}{}{}{} 
% 5 {} token are mandatory
 
  \abstract
   {The CoRoT and {\it Kepler} space-borne missions have provided us with a wealth of high-quality observational data that allows for seismic inferences of stellar interiors. This requires the computation of precise and accurate theoretical frequencies, but imperfect modeling of the uppermost stellar layers introduces systematic errors. To overcome this problem, an empirical correction has been introduced by Kjeldsen et al. (2008, ApJ, 683, L175) and is now commonly used for seismic inferences. Nevertheless, we still lack a physical justification allowing for the quantification of the surface-effect corrections.}
   {Our aim is to constrain the surface-effect corrections across the Hertzsprung-Russell (HR) diagram using a set of 3D hydrodynamical simulations.}
   {We used a grid of these simulations computed with the CO$^5$BOLD code to model the outer layers of solar-like stars. 
Upper layers
of the corresponding 1D standard models  
were
then replaced by the layers obtained from the horizontally averaged 3D models. The frequency differences between these patched models and the 1D standard models 
were 
then calculated using the adiabatic approximation and 
allowed
us to constrain 
the \citeauthor{Kjeldsen08}
power law, as well as a Lorentzian formulation.}
   {We find that the  
surface effects
on modal frequencies depend significantly on both the effective temperature and the surface gravity. We further provide the variation 
in
the parameters related to the surface-effect corrections using 
their
power law as well as a Lorentzian formulation. Scaling relations between these parameters and the elevation (related to the Mach number) is also provided. The Lorentzian formulation is shown to be more robust for the whole frequency spectrum, while the power law is not suitable for the frequency shifts in the frequency range above $\nu_{\rm max}$. Finally, we show that,
owing to turbulent pressure, the elevation of the uppermost layers
modifies the location of the hydrogen 
ionization
zone and consequently introduces glitches in the surface effects for models with high (low) effective temperature (surface gravity).
   }
   {Surface-effect corrections vary significantly across the HR diagram. Therefore, empirical relations 
like those
by Kjeldsen et al. 
must not be calibrated on the Sun but should 
instead
be constrained using realistic physical 
modeling
as provided by 3D hydrodynamical simulations.}


   \maketitle
%
%________________________________________________________________

\section{Introduction}

Our knowledge 
of
solar-like oscillations has been recently improved thanks to the precise observations obtained with the CoRoT \citep{Baglin06ESASP,Baglin06cosp,Michel08} and {\it Kepler} \citep{Borucki10} spacecrafts. Such oscillations are stochastically excited and damped by turbulent motions in the outermost layers of convective regions \citep[for a review, see][]{Samadi11}. It is allowed to perform precise seismic 
determination
of both the global stellar parameters and stellar interiors by the detection of a large number of consecutive radial orders and angular degrees, as well as 
mode identification \citep[for a review, see][]{ChaplinMiglio2013}. A striking example is the ability of asteroseismology to improve the determination of stellar ages \citep[e.g.,][]{Lebreton14,SilvaAguirre2015}.

However, asteroseimology still suffers from uncertainties that prevent us from making the best use of the precise seismic data. Indeed, 
precise and accurate determination of stellar interiors requires that we are able to model solar-like stars as realistically as possible. Although rotation and magnetic fields should affect properties of equilibrium structure and oscillation, we still do not have definitive methods 
of including
them 
in the modeling.
But, even if we neglect them, the deficient 
modeling
of the 
uppermost
layers of stars (hereafter surface effects) are already substantial obstacles to accurately  
determining
oscillation frequencies. 

For the Sun, a systematic discrepancy between observed and computed frequencies of the $p$ modes has been 
emphasized
\citep{CD88,Dziembowski88,CD96,CD97} and related to the poor modeling of the near-surface region. For instance, in classical standard models, the mixing-length theory is used to model convection, and this is only valid for efficient convection in deep interiors. However, convection becomes inefficient in the near-surface regions. Therefore, more complex physical processes 
need to be accounted for, such as compressible turbulence, which is neglected by the mixing-length theory.

For the Sun, there have been many attempts to analyze the surface effects, 
and they are
based on 
a
more sophisticated treatment of the surface convection than the mixing-length theory. Indeed, the frequencies of the high-order $p$-modes were found to be affected by the treatment of convection \citep[e.g.,][]{Brown84, Zhugzhda94, Schlattl97, Petrovay07}. Similar attempts 
have also been made
with more realistic models constructed using 3D hydrodynamical simulations \citep{Stein91, Rosenthal99, Yang07, Piau14, Bhattacharya15}. It has been found that turbulent pressure (neglected in standard models) plays a major role in modifying the frequencies, since the induced elevation of the outer layer leads to the decrease 
in
the frequencies. We also note that the thermal timescale is comparable to the oscillation periods in the near-surface regions. Consequently, one can expect that nonadiabatic effects on frequencies can have a non-negligible impact and should also be taken into account. This has 
been done partly
by several authors \citep[e.g.,][]{Houdek2010,Grigahcene2013}, but given the uncertainties related to the treatment of the convection--pulsation coupling, this would deserve 
being
investigated more thoroughly. 

In the absence of any definitive conclusions on the surface effects, \cite{Kjeldsen08} 
have
proposed an empirical power law to correct the theoretical frequency \citep[see also][]{Ball2014}. They 
provide
the value of the power index in the power law by analyzing the difference between the observed solar frequencies and the theoretical frequencies computed with Model S \citep{CD96}. Since then, many authors have adopted the value to correct the computed frequencies for other stars than the Sun \cite[e.g.,][]{CD10, Dogan10, Gruberbauer13}. Alternative strategies have been adopted to determine the value of the index \citep[e.g.,][]{Gruberbauer12, Lebreton14}, but still adopting the power-law function of \cite{Kjeldsen08}. Indeed, the correction is required since, without it, it is difficult to find a good model from comparison with the observations. At this stage, however, there is no physical justification for constraining the empirical surface-effect corrections. 

In this article, we construct outer-layer models with 3D hydrodynamical simulations including non-local radiation transport for different types of solar-like stars (\S\ref{sec:3Dgrid}). The constructed 3D layers are patched to the standard interior model (\S\ref{sec:PM}). Then, the frequency differences with the standard models are evaluated (\S\ref{sec:adosc}). We discuss the dependence on effective temperature and surface gravity 
when
performing the functional fittings (\S\ref{sec:fit}). Discussion and 
conclusions
are given in Sections \ref{sec:discuss} and \ref{sec:conclusion}, respectively.

\begin{table}
  \caption{Characteristics of the 3D hydrodynamical models.}
  \label{tab:3D}
  \begin{center}
    \begin{tabular}{c|cccc}\hline\hline
      Model & $T_{\rm eff}$ & $\log\,g$ & $T_{\rm b}$        & $\nu_{\rm max}$ \\
            & [K]         &[cm/s$^2$]& [K]               & [$\mu$Hz] \\
      \hline
      A     & 5775        & 4.44     & $1.53\times 10^4$ & 3110 \\
      B     & 6725        & 4.25     & $8.26\times 10^4$ & 1864 \\
      C     & 6486        & 4.00     & $2.84\times 10^4$ & 1067 \\
      D     & 6432        & 4.25     & $2.80\times 10^4$ & 1906 \\
      E     & 6227        & 4.00     & $2.13\times 10^4$ & 1089 \\
      F     & 6102        & 4.25     & $2.53\times 10^4$ & 1956 \\
      G     & 5861        & 4.50     & $2.36\times 10^4$ & 3550 \\
      H     & 5927        & 4.00     & $2.01\times 10^4$ & 1116 \\
      I     & 5885        & 3.50     & $2.02\times 10^4$ & 354.3 \\
      J     & 4969        & 2.50     & $1.46\times 10^4$ & 38.56 \\
      \hline
    \end{tabular}
  \end{center}
  \tablefoot{$T_{\rm eff}$ is the effective temperature, $\log\,g$ the logarithm of the surface gravity, $T_{\rm b}$ the temperature at the bottom of the 3D models, $\nu_{\rm max}$ 
the frequency with maximum amplitude of solar-like oscillations, here estimated by using Eq. (\ref{eq:numax})}
\end{table}

\begin{figure}
  \centering
  \includegraphics[width=\hsize]{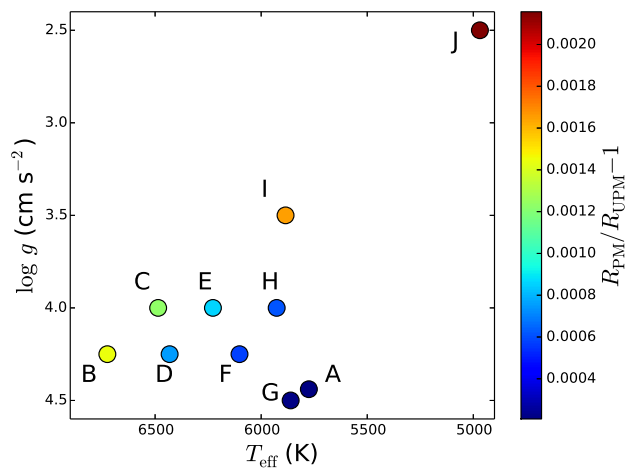}
  \caption{3D hydrodynamical models in the $T_{\rm eff}$-$\log\, g$ plane. The relative difference in the radius between PM and UPM is indicated by the color scale.}
  \label{fig:delRR}
\end{figure}

\begin{figure}
  \centering
  \includegraphics[width=\hsize]{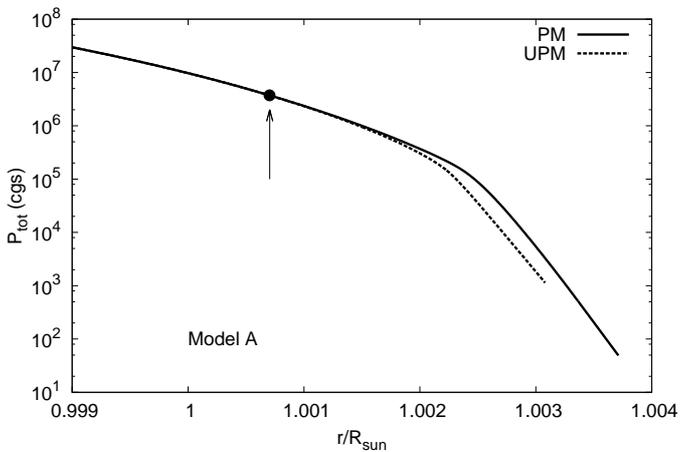}
  \caption{Comparison of the total pressure $P_{\rm tot}$ between PM and UPM of 
Model
A. The abscissa is the radius divided by the solar radius, $r/R_\odot$. The vertical arrow indicates the point of matching between the CESTAM model and the CO$^5$BOLD model. Below the matching point, the profiles of PM and UPM are identical.}
  \label{fig:rp}
\end{figure}

\begin{table*}
  \caption{Characteristics of UPM and PM}
  \label{tab:PSM}
  \centering
  \begin{tabular}{c|ccccccc}
    \hline\hline
    Model & $M$\tablefootmark{{\rm ($a$)}}& Age     & $\alpha_{\rm MLT}$ & $R_{\rm PM}$ & $R_{\rm PM}/R_{\rm UPM}-1$ & stage\tablefootmark{{\rm ($b$)}}\\
          & [$M_\odot$]         & [Gyr]    &                   & [$R_\odot$] &                           & \\ 
    \hline
    A     & 1.01               & 4.61 & 1.65 & 1.00 & $2.08\times 10^{-4}$  & MS\\
    B     & 1.37               & 1.36 & 1.69 & 1.45 & $1.44\times 10^{-3}$  & MS\\
    C     & 1.47               & 2.30 & 1.68 & 2.01 & $1.23\times 10^{-3}$  & MS\\
    D     & 1.26               & 2.60 & 1.69 & 1.39 & $7.54\times 10^{-4}$  & MS\\
    E     & 1.38               & 3.20 & 1.69 & 1.94 & $8.61\times 10^{-4}$  & MS\\
    F     & 1.12               & 5.43 & 1.68 & 1.31 & $5.75\times 10^{-4}$  & MS\\
    G     & 1.08               &0.0209& 1.66 & 0.967& $2.21\times 10^{-4}$  & MS\\
    H     & 1.14               & 6.93 & 1.70 & 1.77 & $6.15\times 10^{-4}$ & SG\\
    I     & 1.73               & 1.76 & 1.65 & 3.83 & $1.65\times 10^{-4}$ & RG\\
    J     & 3.76               &0.213 & 1.61 & 18.2 & $2.16\times 10^{-3}$ & RG\\
    \hline
  \end{tabular}
  \tablefoot{\tablefoottext{a}{The stellar masses for UPM and PM differ by at most a fraction $\sim 10^{-7}$.}
  \tablefoottext{b}{Evolutionary stage. MS, SG, and RG mean the main sequence, subgiant, and 
red giant stages, respectively.}
  }
\end{table*}

\section{Stellar patched models and related eigenfrequencies}

We consider a set of 3D hydrodynamical models for which two types of corresponding 1D models 
were
constructed, the patched models (PMs) and unpatched models (UPMs). Subsequently, the associated frequencies 
were
computed in the framework of the adiabatic approximation.

\subsection{Grid of 3D hydrodynamical models}
\label{sec:3Dgrid}

We use the CO$^5$BOLD code \citep{Freytag12} with the CIFIST grid \citep{Ludwig09}. The adopted chemical mixture is similar to the solar abundances determined by \cite{Asplund05}.
We 
considered
ten models. The global characteristics of the models are summarized in Table \ref{tab:3D} as well as the temperature at the bottom of the 3D models, $T_{\rm b}$, and the frequency of the maximum mode height in the oscillation power spectrum, $\nu_{\rm max}$, estimated by the scaling relation \citep[e.g.,][]{Brown91,Kjeldsen95,Belkacem11}
\begin{eqnarray}
  \frac{\nu_{\rm max}}{\nu_{{\rm max},\odot}}=\frac{g}{g_\odot}\left(\frac{T_{\rm eff}}{T_{{\rm eff},\odot}}\right)^{-1/2},
  \label{eq:numax}
\end{eqnarray}
which arises from the proportionality between $\nu_{\rm max}$ and the cut-off frequency. For solar values, we adopt $\nu_{{\rm max}, \odot} = 3100 \, \mu$Hz, $\log g_\odot=4.438$, and $T_{{\rm eff}, \odot}=$5777 K. 
The location of the selected models in the $T_{\rm eff}$-$\log g$ plane is displayed in Fig. \ref{fig:delRR}.  We note that Model A  corresponds to the Sun.

\subsection{Computation of the patched models}\label{sec:PM}

Following \cite{Trampedach97} and \cite{Samadi07, Samadi08}, 
we construct a corresponding UPM and PM,
for each 3D simulations.
The UPMs 
were
obtained using the CESTAM stellar evolution code \citep{Marques13} by matching the effective temperature, the gravity at the photosphere, and the temperature at the bottom of the 3D layers. The matching 
was
performed through a Levenberg-Marquardt algorithm with three free parameters:
the age, the total mass $M$, and the mixing length parameter $\alpha_{\rm MLT}$. For post-main sequence stages (models H, I, and J), 
the central temperature rather than the age 
was
considered as a free parameter
because of their rapid evolution. 
Convection in the 1D models is treated using the standard 
mixing-length
theory \citep{Vitense58} and thus does not include turbulent pressure. We also 
adopted
the Eddington 
gray 
$T$--$\tau$ relation for the atmosphere of UPMs.

Subsequently, the UPMs and corresponding 3D hydrodynamical models 
were
matched to obtain the PMs. To this end, temporal and horizontal averages 
were
performed at constant geometrical depth for the 3D models. Figure~\ref{fig:rp} compares the profiles of the total pressure in the PM and UPM of 
Model
A. They are, by construction, identical below the matching point. Above, the pressure of the UPM is 
lower
than of the PM at the same radius (which denotes the distance from the center). The difference in the pressure stratifications mainly depends on the effect of turbulent pressure in the PM. This additional source of pressure modifies the hydrostatic equilibrium so that, at the photosphere, the radii of the PM ($R_{\rm PM}$) and UPM ($R_{\rm UPM}$) are different. As 
summarized
in Table \ref{tab:PSM}, the relative difference $R_{\rm PM}/R_{\rm UPM}-1$ ranges between 0.01 and 0.2\%. It increases with increasing effective temperature or with decreasing surface gravity (see Fig. \ref{fig:delRR}). This can be explained as follows: With increasing effective temperature, the larger H$^-$ opacity requires a steeper temperature gradient for ensuring the required energy transfer. Therefore, the convective velocity increases. With decreasing surface gravity, the density in the near-surface region decreases. For ensuring the required convective transport, the convective velocity should be higher. In both cases, the contribution of turbulent pressure becomes prominent.

\begin{figure}[th!]
  \centering
  \includegraphics[width=\hsize]{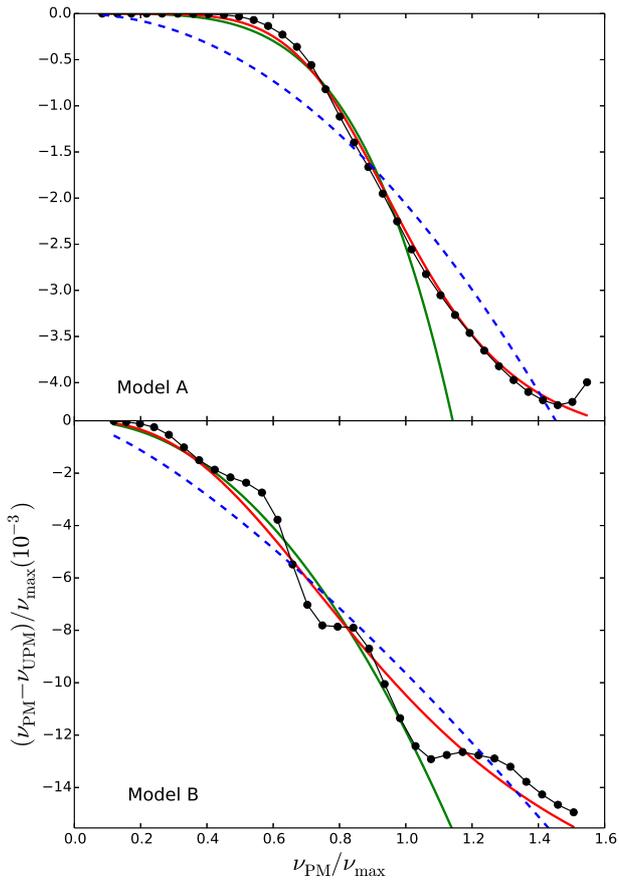}
  \caption{Frequency differences of radial modes between PM and UPM divided by $\nu_{\rm max}$ as a function of $\nu_{\rm PM}/\nu_{\rm max}$ (black solid line with dots). The green full line and the blue dashed line are least-square fittings with 
the \cite{Kjeldsen08}
power law, $a(\nu_{\rm PM}/\nu_{\rm max})^b$ in the range $0<\nu/\nu_{\rm max}<1.05$ and in the whole range, respectively. The red line shows the modified Lorentzian, $\alpha[1-1/\{1+(\nu_{\rm PM}/\nu_{\rm max})^\beta\}]$.}
  \label{fig:dnu}
\end{figure}

\subsection{Computation of adiabatic oscillations}\label{sec:adosc}

We 
computed
adiabatic oscillations using the ADIPLS code \citep{CD08} for both the PMs and UPMs. We 
considered
only radial modes below the cut-off frequency. 
For the PMs, the main 
problem
is to consider the turbulent pressure in 
computing
the frequencies. A fully non-adiabatic oscillation code including a time-dependent treatment of convection is a natural way to account for the turbulent pressure and its perturbation. However, for 
the
sake of simplicity, we 
adopted
the adiabatic approach as proposed by \cite{Rosenthal99}, who 
considered two simplifying approximations that allow us to use a classical adiabatic oscillation code in a simple way:
 \begin{itemize}
\item the gas $\Gamma_1$ approximation, for which the relative Lagrangian perturbation of turbulent pressure equals the Lagrangian perturbation of gas pressure,
\begin{eqnarray}
\label{gas_gamma}
  \frac{\delta P_{\rm turb}}{P_{\rm turb}}\simeq\frac{\delta P_{\rm g}}{P_{\rm g}} = \Gamma_1 \frac{\delta \rho}{\rho},
\end{eqnarray}
where $\delta$ denotes the Lagrangian perturbation, $P_{\rm turb}$ the turbulent pressure, and $P_{\rm g}$ the gas pressure. Equation~(\ref{gas_gamma}) permits us to express the perturbation of turbulent pressure as a function of the density perturbation. 
\item the reduced $\Gamma_1$ approximation, for which $\Gamma_1$ is modified to become $\Gamma^r_1\equiv (P_{\rm g}/P_{\rm tot})\;\Gamma_1$. It implies that the perturbation of turbulent pressure is neglected so that 
\begin{eqnarray}
  \frac{\delta P_{\rm turb}}{P_{\rm turb}}=0.
\end{eqnarray}
\end{itemize}

By comparing 
these
two  
approximations, \cite{Rosenthal99} found that the frequencies computed with the gas $\Gamma_1$ are closer to the observed solar frequencies. Therefore, we  
have adopted
the gas $\Gamma_1$ treatment in this study. As mentioned above, however, a consistent way to consider turbulent pressure is to compute nonadiabatic oscillations. We will pursue the nonadiabatic case in 
our next
paper.

Figure~\ref{fig:dnu} shows the difference 
in
radial-mode frequencies between the PM and UPM for 
Models
A and B. The PM has a larger radius than the UPM due to the elevation by the turbulent pressure. The cavity where acoustic waves propagate then becomes larger, and the mode frequencies become lower for PM. As a consequence, the difference is negative for all of the models. We can also see that the difference becomes 
greater
with increasing frequency. As the frequency becomes higher, the mode propagates 
farther
in the outer region, so that it is affected by the surface effects more strongly. 

For several models, the profiles of the frequency difference show oscillatory features as found in the bottom panel of Fig. \ref{fig:dnu}. Actually, they are caused by the acoustic glitch in the hydrogen ionization zone and by the difference in the size of the acoustic cavity due to the turbulence elevation. 
This
detail is discussed in section \ref{sec:glitch}.   

\section{Functional fittings to frequency difference}
\label{sec:fit}

The power law proposed by \cite{Kjeldsen08} is commonly used for correcting theoretical oscillation frequencies and for reproducing the observed frequencies. It is then worth discussing the validity of the fitting with the power law using the frequency difference between the PM and UPM. As mentioned below, however, the power law is not suitable at high frequency. We also propose a formulation based on a Lorentzian, which reduces to the power law in the 
low-frequency
limit.

\begin{table}
  \caption{Reduced values of the squared deviation, ${\cal D}^{\rm r}$}
  \label{tab:err}
  \centering
  \begin{tabular}{c|ccc}\hline\hline
    Model & \multicolumn{2}{c}{Power law} & Lorentzian \\ 
          & whole range & $0<\nu/\nu_{\rm max}<1.05$ &            \\  
    \hline
    A     & 1.61$\times 10^{-7}$ & 8.16$\times 10^{-9}$ & 6.58$\times 10^{-9}$ \\
    B     & 1.54$\times 10^{-6}$ & 3.18$\times 10^{-7}$ & 4.86$\times 10^{-7}$ \\
    C     & 1.94$\times 10^{-6}$ & 4.30$\times 10^{-7}$ & 6.60$\times 10^{-7}$ \\
    D     & 1.11$\times 10^{-6}$ & 1.47$\times 10^{-7}$ & 2.90$\times 10^{-7}$ \\
    E     & 1.57$\times 10^{-6}$ & 2.59$\times 10^{-7}$ & 4.03$\times 10^{-7}$ \\
    F     & 7.86$\times 10^{-7}$ & 6.41$\times 10^{-8}$ & 1.62$\times 10^{-7}$ \\
    G     & 1.94$\times 10^{-7}$ & 8.45$\times 10^{-10}$ & 2.95$\times 10^{-8}$ \\ 
    H     & 8.73$\times 10^{-7}$ & 1.03$\times 10^{-7}$ & 1.77$\times 10^{-7}$ \\
    I     & 3.69$\times 10^{-6}$ & 1.11$\times 10^{-6}$ & 1.06$\times 10^{-6}$ \\
    J     & 1.67$\times 10^{-6}$ & 5.09$\times 10^{-7}$ & 6.61$\times 10^{-7}$ \\
    \hline
  \end{tabular}
  \tablefoot{Concerning the power law, the deviation in the whole frequency range is also shown for comparison (the second column).}
\end{table}

\subsection{The \cite{Kjeldsen08} power law}
\label{sec:k08}

Since the PMs include realistic profiles of the upper stellar atmosphere, we 
considered 
their oscillation frequencies 
to be 
the observed ones. The correction proposed by \cite{Kjeldsen08} thus becomes
\begin{eqnarray}
  \frac{\delta\nu}{\nu_{\rm max}}=a\left(\frac{\nu_{\rm PM}}{\nu_{\rm max}}\right)^b,
  \label{eq:K08}
\end{eqnarray} 
where $\delta \nu$ is correction of the frequency corresponding to $\nu_{\rm PM}-\nu_{\rm UPM}$, and the coefficients $a$ and $b$ are free parameters. 
Parameter
$a$ is, in the present study, non-dimensional. On the other hand, the parameter had a dimension in \cite{Kjeldsen08}, since the frequency correction $\delta\nu$ 
on the lefthand
side was not divided by $\nu_{\rm max}$. By definition, $a$ is thus the value of $\delta\nu/\nu_{\rm max}$ at $\nu_{\rm PM}=\nu_{\rm max}$.

\subsection{Parameter fitting}
\label{sec:param_fit}

To determine the parameters $a$ and $b$, we 
performed
a least-square analysis. We fit Eq.~(\ref{eq:K08}) to the computed frequency difference so that the summation of the squared deviation,
\begin{eqnarray}
  {\cal D}=\sum^{N}_{i}\left[\frac{\nu_{{\rm PM},i} - \nu_{{\rm UPM},i} - \delta\nu_i}{\nu_{\rm max}}\right]^{2},
  \label{eq:D}
\end{eqnarray} 
is minimized, where $i$ denotes the label of the eigenmodes, and $N$ is the number of the modes. 
Figure \ref{fig:dnu} shows the results for Models A and B 
They are determined in the range $0<\nu_{\rm PM}/\nu_{\rm max}<1.05$. Table \ref{tab:err} shows the reduced values of Eq. (\ref{eq:D}), i.e., ${\cal D}^{\rm r}(\equiv {\cal D}/N)$. For comparison, we also 
considered
the deviation of the fit performed in the whole frequency range. The reduced values of the deviation in the whole range are significantly larger, which clearly shows that the power law is not suitable for the high frequency range (dashed blue curve in Fig. \ref{fig:dnu}). 
However,
the power law was originally used for fitting 
intermediate-order
modes of 
only the Sun,
and the  
high-frequency
modes were not taken into account in \cite{Kjeldsen08}. 

The values of $a$ and $b$ depend strongly on the choice of the fitting range. This point was also 
made
by \cite{Kjeldsen08} and is discussed in Sect.~\ref{sec:fitrange}.

\subsubsection{Parameters across the $T_{\rm eff}$--$g$ plane}

\begin{figure}
  \centering
  \includegraphics[width=\hsize]{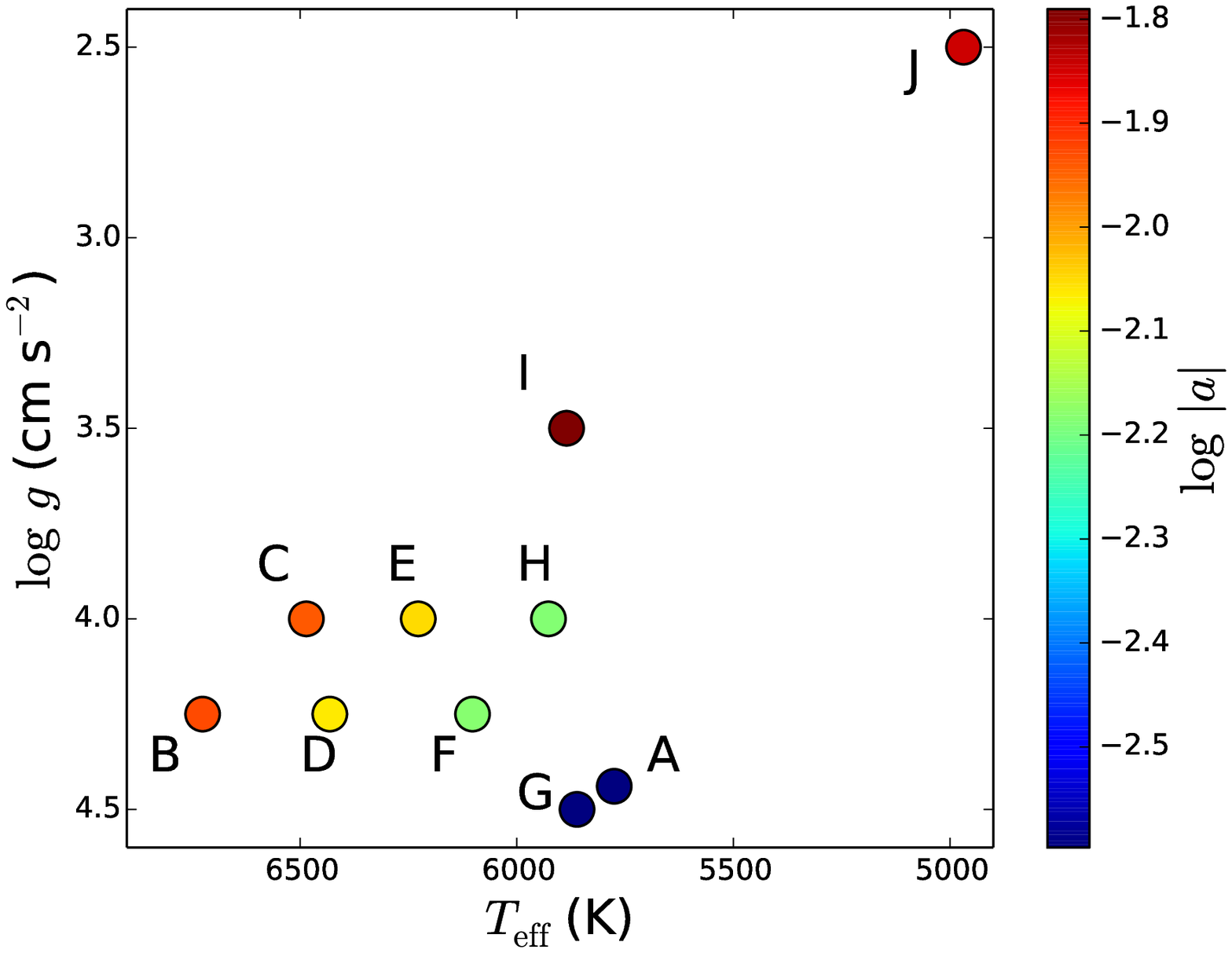}
  \includegraphics[width=\hsize]{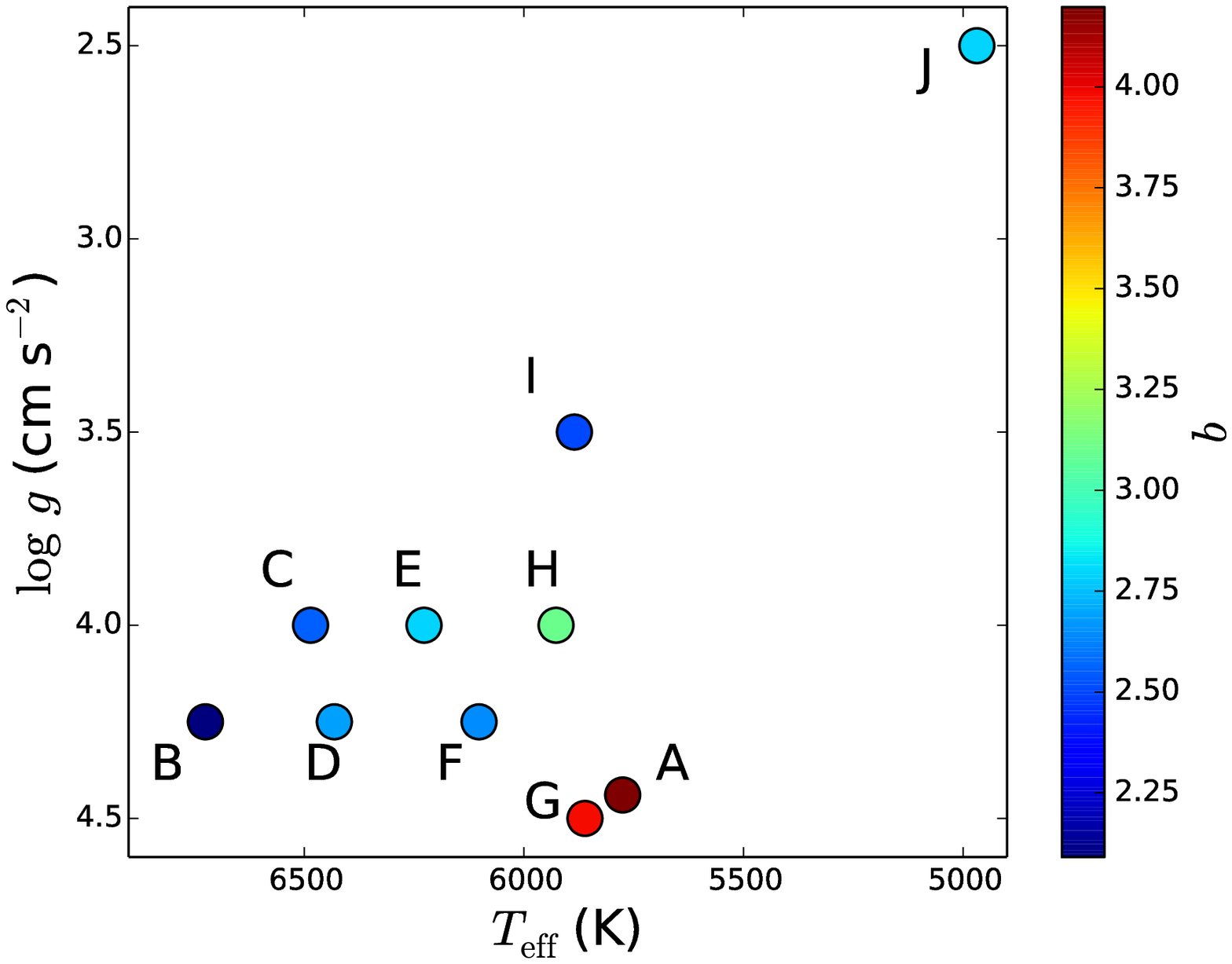}
  \caption{
Values 
of $\log |a|$ (top) and $b$ (bottom) obtained by the fitting with Eq.(\ref{eq:K08}) in the range $0<\nu/\nu_{\rm max}<1.05$ are indicated by 
colors 
on the $T_{\rm eff}$--$\log g$ plane.}
  \label{fig:Teff-g-ab}
\end{figure}

Regardless of the fitting range, we have found that the power index $b$ certainly varies with models, as shown in Fig. \ref{fig:Teff-g-ab} and Table \ref{tab:res}. \cite{Kjeldsen08} 
propose
$b=4.90$ in the case of the Sun, and many studies have adopted this value for other stars \citep[e.g.,][]{CD10, Dogan10, Metcalfe10, Tang11, Brandao11, VanEylen12, Dogan13, Gilliland13, Gruberbauer13}. When $b$ is fixed, a simple fit provides the value of $a$ \citep{Kjeldsen08}.

Then, a least-square fit provides the variation 
in
$a$ and $b$ as a function of $\log T_{\rm eff}$ and $\log\,g$, according to
\begin{eqnarray}
  && \log |a| = 8.13\log T_{\rm eff} - 0.670\log g - 30.2, \label{eq:a}\\
  && \log b = -3.16\log T_{\rm eff} + 0.184\log g + 11.7. \label{eq:b}
\end{eqnarray}
To measure the deviations of the computed coefficients from the above equations, we define
\begin{eqnarray}
  \sigma_f = \sqrt{\frac{1}{N-1}\sum_i(f_i - \tilde{f}_i)^2},
  \label{eq:SD}
\end{eqnarray}
where $f$ corresponds to $a$ or $b$ (not $\log |a|$ or $\log b$), $i$ is a label indicating each model, $N$ 
the number of the models, and $\tilde{f}$ 
the value obtained by Eq. (\ref{eq:a}) or (\ref{eq:b}). Here we have $N=10$. We obtain $\sigma_a=4.6\times 10^{-4}$ and $\sigma_b=9.1\times 10^{-2}$, which means that the relative deviations are 
on 
the order of 10\,\%.

The trends of $b$ and of the absolute value of $a$ are 
the
opposite. With increasing $T_{\rm eff}$ or with decreasing $g$, $b$ and $|a|$ become smaller and larger, respectively. We recall that the coefficient $a$ is the value of $\delta\nu/\nu_{\rm max}$ at $\nu=\nu_{\rm max}$ by definition. Thus, it is the representative scale for $\delta\nu/\nu_{\rm max}$. \cite{Rosenthal99} suggested that this scale should be proportional to the elevation of the radius. Although their analysis 
was 
limited to the Sun, we confirm this tendency among the different models of stars. 
The trend of coefficient $b$
is mainly determined by the low-order modes and we find that it also scales with the elevation. We discuss this issue in detail in Sect.~\ref{sec:discuss_ab}.

\begin{table}
  \caption{Coefficients of the \cite{Kjeldsen08} and the Lorentzian fittings.}
  \label{tab:res}
  \centering
  \begin{tabular}{c|cc|cc}
    \hline\hline
          &\multicolumn{2}{c|}{\citeauthor{Kjeldsen08}} & \multicolumn{2}{c}{Lorentzian} \\
    Model & $a$                  & $b$  & $\alpha/2$              & $\beta$ \\
    \hline
    A     & $-2.53\times 10^{-3}$ & 4.20 & $-2.36\times 10^{-3}$ & 5.66 \\
    B     & $-1.18\times 10^{-2}$ & 2.09 & $-1.05\times 10^{-2}$ & 2.56 \\
    C     & $-1.15\times 10^{-2}$ & 2.55 & $-9.65\times 10^{-3}$ & 2.93 \\
    D     & $-8.63\times 10^{-3}$ & 2.69 & $-7.13\times 10^{-3}$ & 3.03 \\
    E     & $-8.89\times 10^{-3}$ & 2.79 & $-7.53\times 10^{-3}$ & 3.26 \\
    F     & $-6.54\times 10^{-3}$ & 2.64 & $-5.86\times 10^{-3}$ & 3.26 \\
    G     & $-2.54\times 10^{-3}$ & 3.98 & $-2.33\times 10^{-3}$ & 5.21 \\
    H     & $-6.50\times 10^{-3}$ & 3.10 & $-5.70\times 10^{-3}$ & 3.74 \\
    I     & $-1.62\times 10^{-2}$ & 2.51 & $-1.35\times 10^{-2}$ & 2.93 \\
    J     & $-1.43\times 10^{-2}$ & 2.79 & $-1.18\times 10^{-2}$ & 3.27 \\ 
    \hline
  \end{tabular}
\end{table}

\subsection{Fitting a modified Lorentzian}
\label{sec:Lorentzian}
\begin{figure}
  \centering
  \includegraphics[width=\hsize]{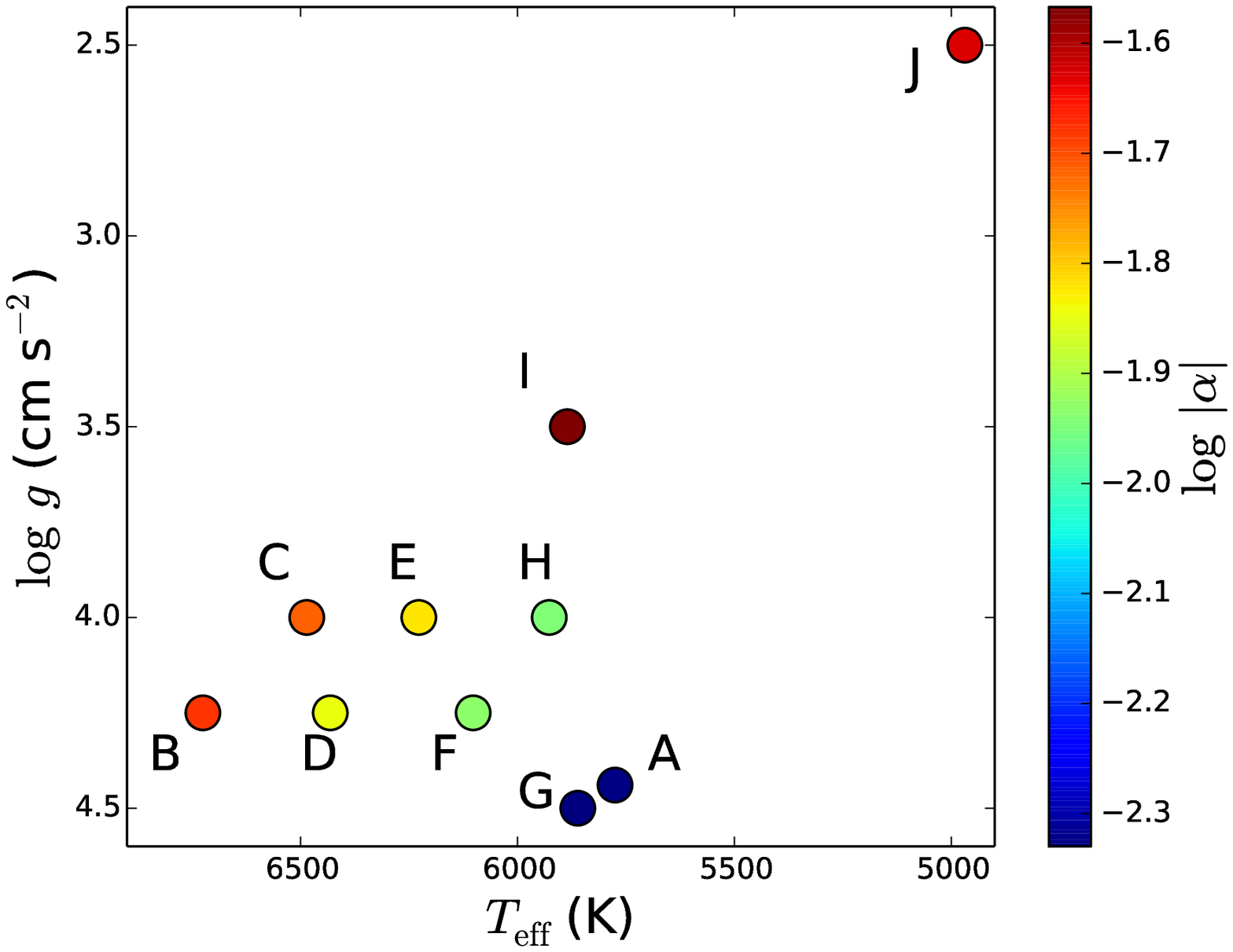}
  \includegraphics[width=\hsize]{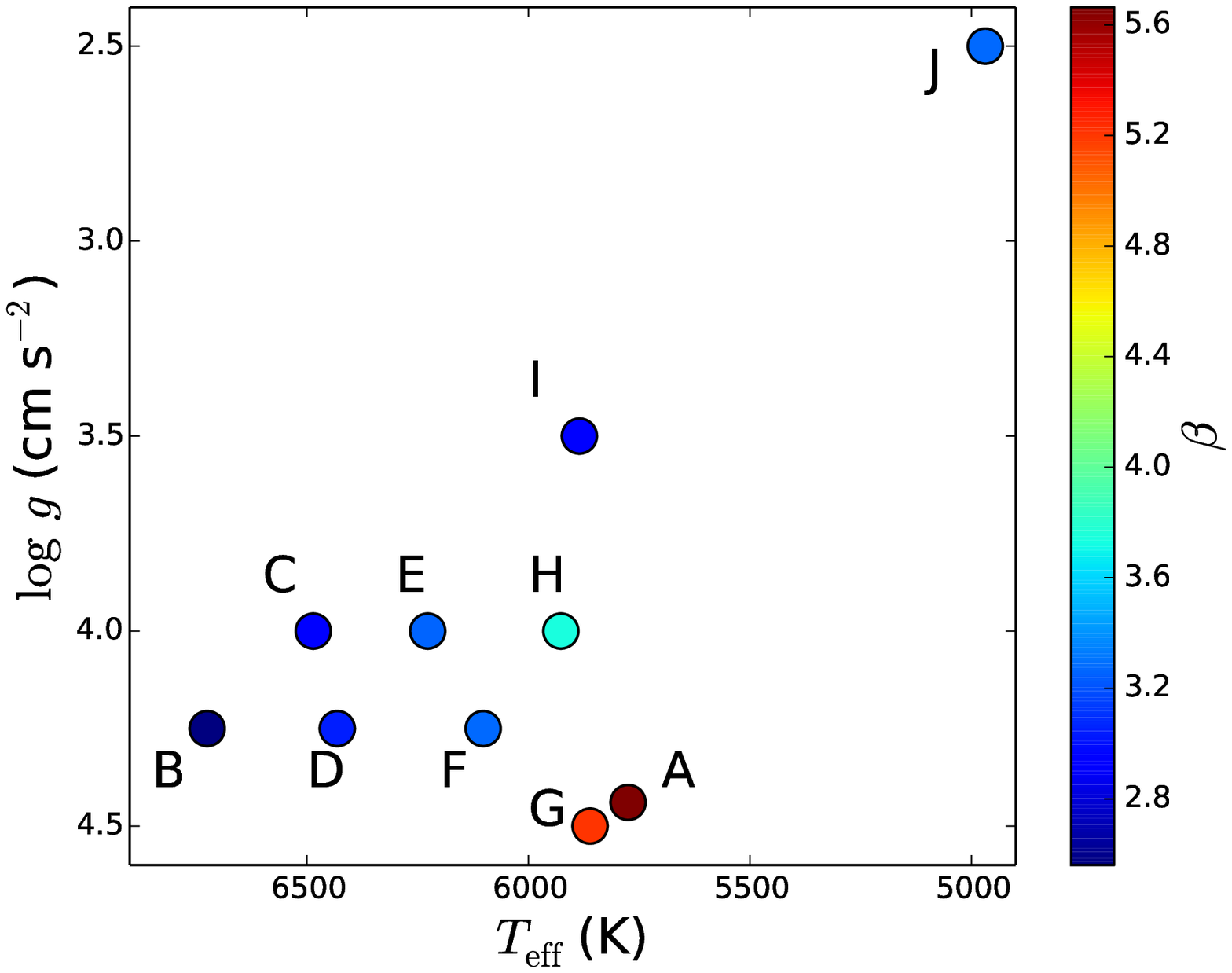}
  \caption{
Values 
of $\log |\alpha|$ (top) and $\beta$ (bottom) obtained by the fitting with Eq. (\ref{eq:Lorentzian}) are indicated by color on the $T_{\rm eff}$--$\log g$ plane.}
  \label{fig:Teff-g-alphabeta}
\end{figure}

As mentioned in Sect.~\ref{sec:k08}, the power-law fitting does not work in the high-frequency range, since the gradient of the frequency difference becomes smaller. \cite{Ball2014} 
propose 
a correction functional with mode inertia and 
demonstrate
that it can be well 
fit 
to the differences between the BiSON frequencies \citep{Broomhall09} and the standard model frequencies in the whole frequency range. For our models, we find that it works well in limited frequency ranges, but not in the whole range. Then, we alternatively propose a formulation based on a modified Lorentzian function, 
\begin{eqnarray}
  \frac{\delta\nu}{\nu_{\rm max}}=\alpha\left[1-\frac{1}{1+(\nu_{\rm PM}/\nu_{\rm max})^\beta}\right],
  \label{eq:Lorentzian}
\end{eqnarray}
where the first term in the brackets ensures that $\delta \nu$ 
becomes 
zero when $\nu_{\rm PM}=0$. By this definition, $\alpha/2$ is the value of $\delta\nu/\nu_{\rm max}$ at $\nu_{\rm PM}=\nu_{\rm max}$. As for $a$ and $b$, the coefficients $\alpha$ and $\beta$ are determined with a least-square analysis. The results are shown in Fig. \ref{fig:Teff-g-alphabeta} and Table \ref{tab:res}.

Figure \ref{fig:dnu} compares the fittings obtained using either Eq. (\ref{eq:K08}) or (\ref{eq:Lorentzian}). The Lorentzian profile is found to fit the whole frequency range. We note that, as shown by Table.~\ref{tab:err}, when we consider the frequency range $0<\nu/\nu_{\rm max}<1.05$, both the power law and the Lorentzian profile provide us with the same magnitude for the error. Indeed, this is 
because
Eq.~(\ref{eq:Lorentzian}) reduces to Eq.~(\ref{eq:K08}) in the low-frequency limit. In addition, it is worth mentioning that the parameters of the Lorentzian profile are weakly sensitive to the glitches in the intermediate frequency range. Indeed, they are mainly determined by the low- and high-frequency limits.

Similarly to the power-law fitting, we derive the variation 
in 
$\alpha$ and $\beta$ with $\log T_{\rm eff}$ and $\log\,g$,
\begin{eqnarray}
  &&\log |\alpha|=7.69\log T_{\rm eff}-0.629\log g-28.5, \label{eq:alpha}\\
  &&\log\beta=-3.86\log T_{\rm eff}+0.235\log g+14.2. \label{eq:beta}
\end{eqnarray}
The deviations evaluated using Eq. (\ref{eq:SD}) are $\sigma_\alpha=5.0\times 10^{-2}$ and $\sigma_\beta=3.1\times 10^{-2}$. The deviation of $\alpha$ is much larger compared to the case of $a$, while those of $\beta$ and $b$ are of the same order. By definition, the trends 
in 
$a$ and $\alpha$ on $\log T_{\rm eff}$ and $\log g$ must be similar. The trend 
in 
$\beta$ is also found to be similar to that 
in 
$b$.

\section{Discussion}\label{sec:discuss}

\subsection{Scalings with the atmosphere elevation}
\label{sec:discuss_ab}

As discussed in Sect.~\ref{sec:fit}, the absolute values of 
coefficients $a$ and $\alpha$ increase with decreasing surface gravity or increasing effective temperature, while 
coefficients $b$ and $\beta$ show the opposite trends. Here, we show that these trends are related to the elevation of the outer layers 
because of 
the turbulent pressure in PM.

\subsubsection{Coefficients $a$ and $\alpha$}

\begin{figure}
  \centering
  \includegraphics[width=\hsize]{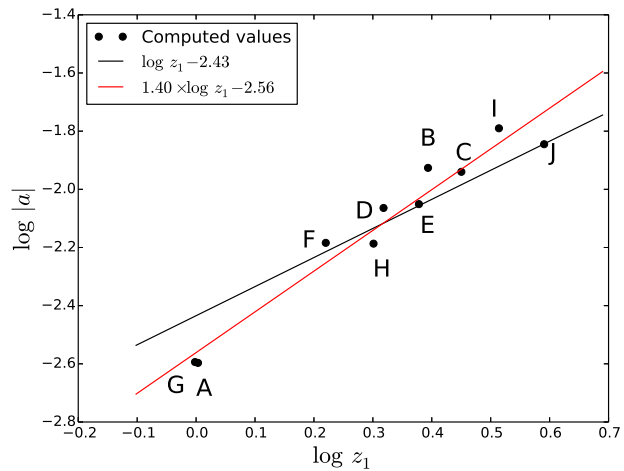}
  \caption{Plots of the logarithm of $|a|$ and the fitting line as a function of $\log z_1$. The black line is the result of the least-square fitting fixing the power-law index of $z_1$ to unity, namely, 
corresponding
to Eq. (\ref{eq:z1_Teffg}), $|a|\propto z_1$. For the red line, the power-law index is also set as the free parameter of the fitting.}
  \label{fig:z1-a}
\end{figure}

According to \cite{Rosenthal99}, the frequency change between two similar models can be estimated as
\begin{eqnarray}
  \frac{\delta\nu}{\nu}\simeq\frac{2\Delta\nu}{c_{\rm ph}}\Delta R,
  \label{eq:Rosenthal99}
\end{eqnarray}
where $\Delta R$ is the difference of radius between the two models, $c$ 
the sound speed, and $c_{\rm ph}$ 
its value at the photosphere. 
Here,
$\Delta\nu$ is the asymptotic large separation,
\begin{eqnarray}
  \Delta\nu\equiv\frac{1}{2}\left(\int^R_0\frac{\diff r}{c}\right)^{-1}.
\end{eqnarray} 
Given 
that $c_{\rm ph}$ scales as $\sqrt{T_{\rm eff}}$, we simplify the scaling relation of Eq. (\ref{eq:Rosenthal99}) as
\begin{eqnarray} 
  \frac{\delta\nu}{\nu}\propto z_1\equiv\frac{\Delta R}{\Delta R_\odot}\left(\frac{T_{\rm eff}}{T_{{\rm eff},\odot}}\right)^{-1/2}\frac{\Delta\nu}{\Delta\nu_{\odot}}.
  \label{eq:z1}
\end{eqnarray}

Here, we consider $\Delta R$ as the difference of the radius between PM and UPM, $R_{\rm PM}-R_{\rm UPM}$. To derive a 
scale 
for the latter quantity, we start by dealing with the ratio of the turbulent pressure to the total pressure $\gamma\equiv P_{\rm turb}/P_{\rm tot}$. Hydrostatic equilibrium implies
\begin{eqnarray}
  \diff\ln P_{\rm tot}=\frac{\diff r}{H_p},
\end{eqnarray} 
where $H_p=P_{\rm tot}/(\rho g)$ is the pressure scale height. Assuming $\gamma\ll 1$, $\Delta R=\gamma H_p$, and since $H_p$ scales as $T_{\rm eff}/g$, we finally derive
\begin{eqnarray}
  \Delta R\propto z_2\equiv\frac{T_{\rm eff}}{T_{\rm eff, \odot}}\frac{g_\odot}{g}\frac{\gamma}{\gamma_\odot},
\end{eqnarray}
where $\gamma$ is evaluated at the photosphere.

The ratio $\gamma$ depends on the Mach number (defined as ${\cal M}_a\equiv w_{\rm rms}/c$, where $w_{\rm rms}$ is the root mean square of the vertical component of the velocity) because by definition, $P_{\rm turb}=\rho w^2_{\rm rms}$. 
Assuming 
a perfect gas, we obtain $P_{\rm g}\propto\rho T$. Moreover, $c\propto\sqrt{T}$, so that $P_{\rm turb}/P_{\rm g}\propto {\cal M}_a$. Finally, since $\gamma\ll 1$, we establish the scaling relation $\gamma\propto {\cal M}^2_a$. As shown in \cite{Samadi13}, ${\cal M}_a$ scales approximately as
\begin{eqnarray}
  {\cal M}_a\propto z_3\equiv\left(\frac{T_{\rm eff}}{T_{\rm eff, \odot}}\right)^p
  \left(\frac{g}{g_\odot}\right)^{-q},
\end{eqnarray}
where $p=2.35$ and $q=0.152$. Using this scaling relation, we finally have
\begin{eqnarray}
  && z_2=\left(\frac{T_{\rm eff}}{T_{\rm eff, \odot}}\right)^{2p+1}\left(\frac{g}{g_\odot}\right)^{-2q-1}, \\
  && z_1=\left(\frac{T_{\rm eff}}{T_{\rm eff, \odot}}\right)^{2p+0.5}\left(\frac{g}{g_\odot}\right)^{-2q-1}
  \frac{\Delta\nu}{\Delta\nu_\odot}.
  \label{eq:z1_Teffg}
\end{eqnarray}

Figure~\ref{fig:z1-a} shows the value of $a$ as a function of $z_1$, which 
was 
evaluated with Eq. (\ref{eq:z1_Teffg}). For $\Delta\nu$, we 
used 
the average value of the large separation and 
adopted 
$\Delta\nu_\odot=134.9\,\mu$Hz \citep{Kjeldsen95}. We 
carried 
out a least-square fitting, fixing the power index of $z_1$ to unity, or 
leaving 
it free. Although the analytical result gives $|a|\propto z_1$, we obtain $|a|\propto z^{1.40}_1$ and $|\alpha|\propto z^{1.31}_1$ when the power index is free. The deviation from a linear scaling with $z_1$ is likely due to the various simplifications adopted in the derivation of this scaling. As the alternative formulations for Eqs.~(\ref{eq:a}) and (\ref{eq:alpha}), we finally have
\begin{eqnarray}
  && \log |a|     = 7.28\log T_{\rm eff} - 1.83\log g \nonumber \\
  &&\qquad\qquad + 1.40\log\Delta\nu [\mu {\rm Hz}] - 24.8, \\
  && \log |\alpha|= 6.83\log T_{\rm eff} - 1.71\log g \nonumber \\ 
  &&\qquad\qquad + 1.40\log\Delta\nu [\mu {\rm Hz}] - 23.5.
\end{eqnarray}
The deviations defined by Eq. (\ref{eq:SD}) are $\sigma_a=1.8\times 10^{-3}$ and $\sigma_\alpha=3.0\times 10^{-3}$. Then, the above equations are less (more) suitable for $a$ ($\alpha$) than Eq. (\ref{eq:a}) (Eq. (\ref{eq:alpha})). But they are 
at least 
based on physical considerations.

\subsubsection{Power indices $b$ and $\beta$} \label{sec:bbeta}

\begin{figure}
  \centering
  \includegraphics[width=\hsize]{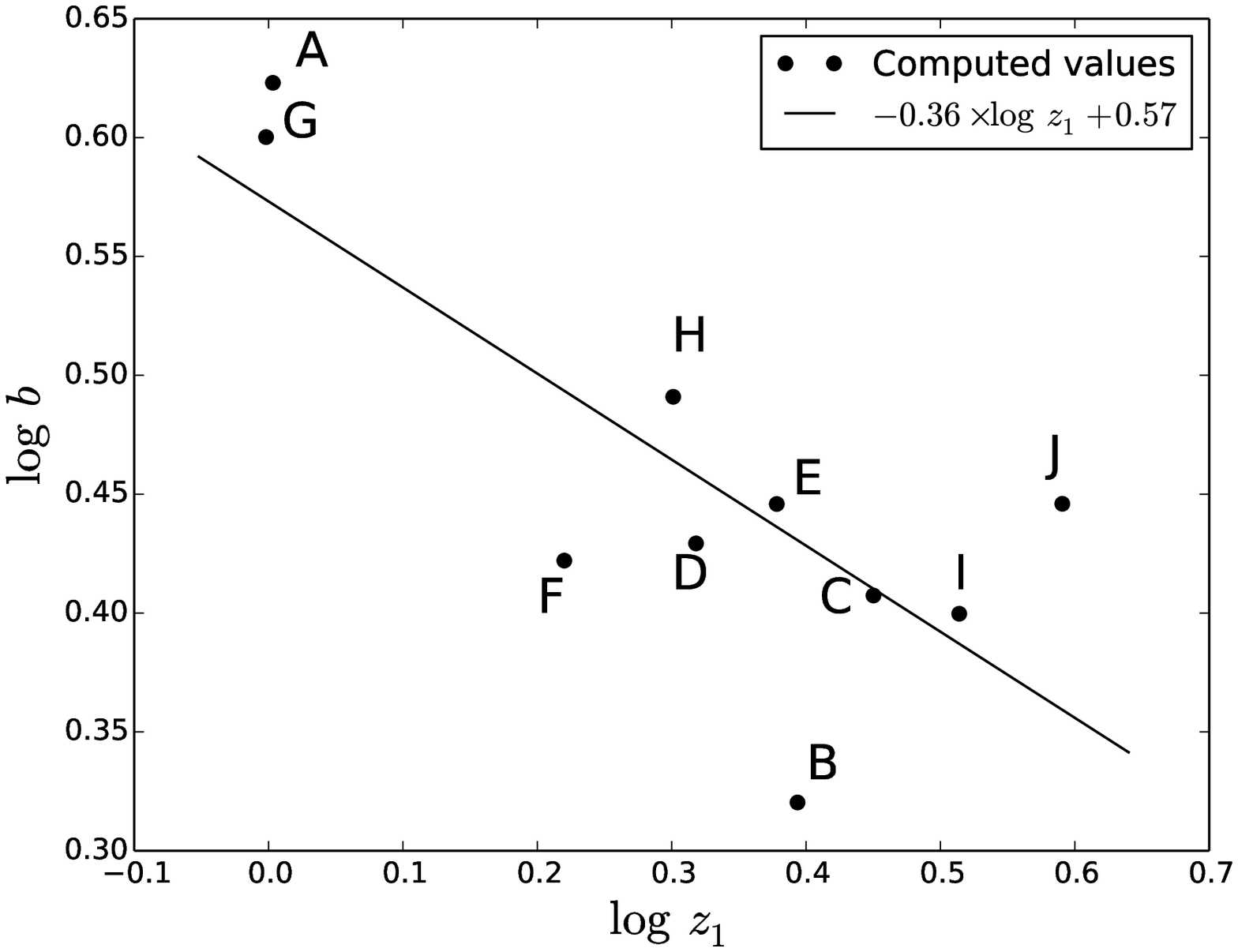}
  \includegraphics[width=\hsize]{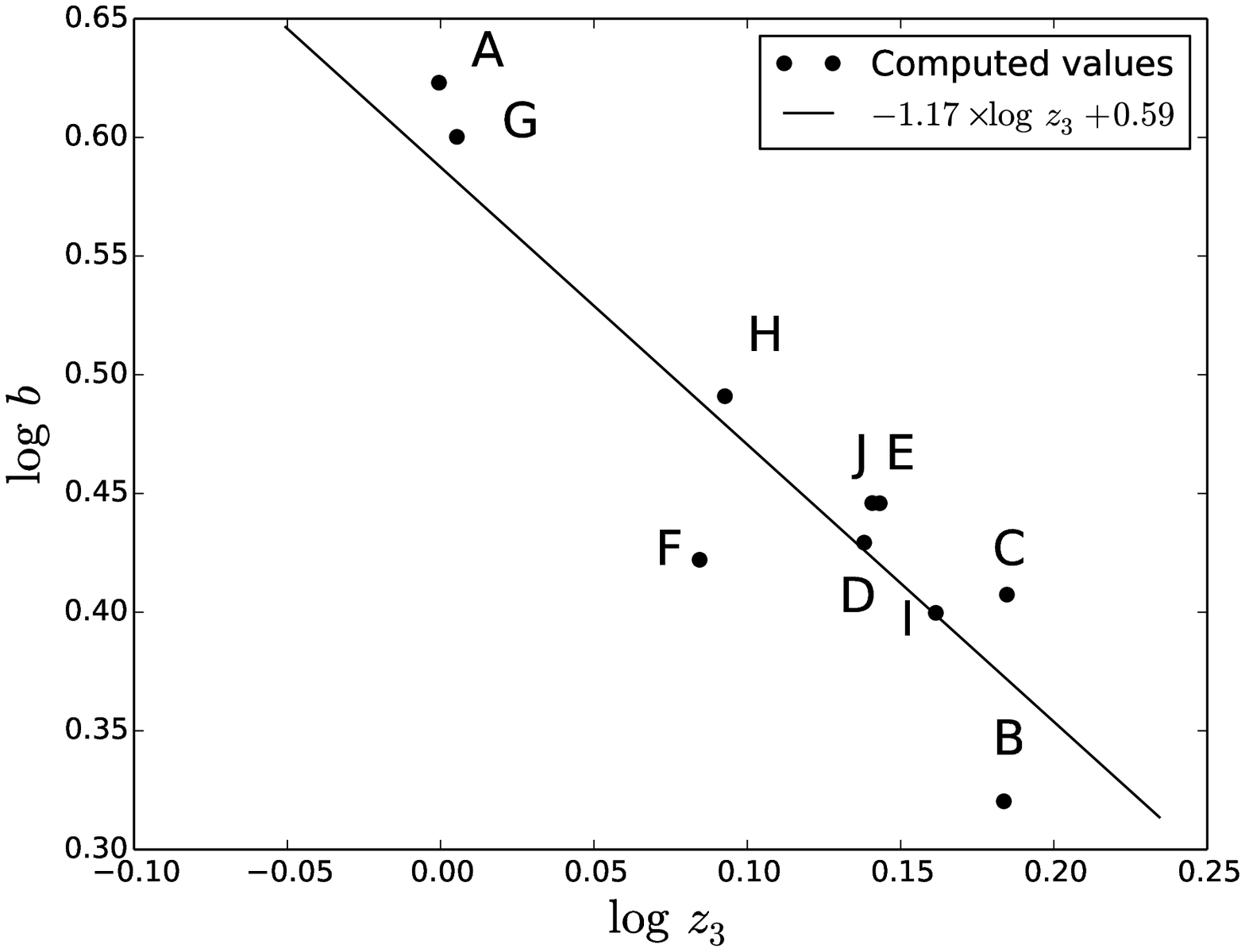}
  \caption{$\log\,b$ as a function of $\log\,z_1$ (top) or of $\log\,z_3$ (bottom). The black line is a result of a least-square fit.}
  \label{fig:z13-b}
\end{figure}

The trend of the coefficients $b$ and $\beta$ on the $T_{\rm eff}$--$\log g$ plane can be also explained in terms of the elevation due to the turbulent pressure. As shown in the top panel of Fig. \ref{fig:z13-b}, both coefficients decrease with increasing $z_1$, although the dispersion is large ($\sigma_b=3.6\times 10^{-1}$, $\sigma_\beta=4.9\times 10^{-1}$). As shown in the bottom panel of Fig. \ref{fig:z13-b}, 
the correlation is better with $z_3$ 
for $b$ and $\beta$
($\sigma_b=2.3\times 10^{-1}$, $\sigma_\beta=3.2\times 10^{-1}$). Using the result of the fitting on the $\log\, z_3$--$\log\, b$ plane, we obtain the physically derived formulations for $b$ and $\beta$,
\begin{eqnarray}
  &&\log\, b   = -2.75\log T_{\rm eff}+0.178\log g+ 10.1, \\
  &&\log\,\beta= -3.37\log T_{\rm eff}+0.218\log g+ 12.4,
\end{eqnarray}
although, in this case, the dispersion for both of them is 
greater 
than for Eqs. (\ref{eq:b}) and (\ref{eq:beta}).

The 
greater
the elevation, the larger the difference in the structure of the convection zone between PM and UPM. Therefore, the low-order modes are more affected by surface effects (see Fig. \ref{fig:dnu}). Consequently, the gradient of $\delta \nu$ at intermediate frequencies becomes smaller and so do the power indices $b$ and $\beta$.

\begin{figure}
  \centering
  \includegraphics[width=\hsize]{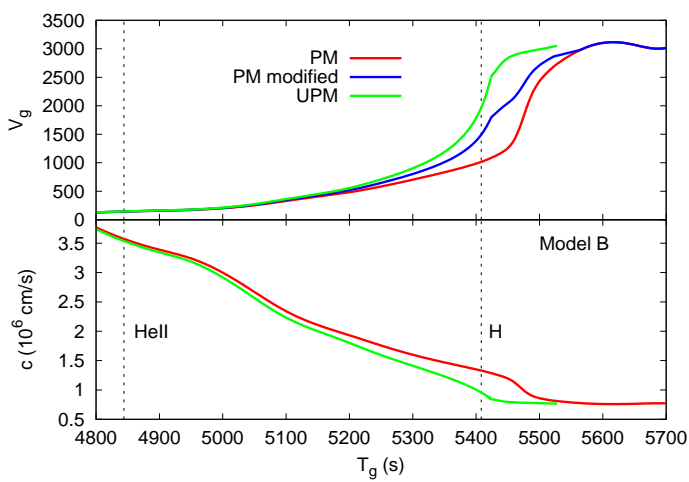}
  \caption{$V_g$ (top) and sound speed $c$ (bottom) in PM and UPM of  
Model
B as a function of the acoustic radius, $T_g(\equiv\int_0^rdr'/c)$. The blue curve in the top panel is the profile for the modified PM, in which the value of $V_g$ is changed to the mean between PM and UPM. The dotted vertical lines indicate the locations of the $\Gamma_1$ bumps due to the HeII and the H ionizations, respectively.}
  \label{fig:Tg-Vgcs}
\end{figure}

\begin{figure}
  \centering
  \includegraphics[width=\hsize]{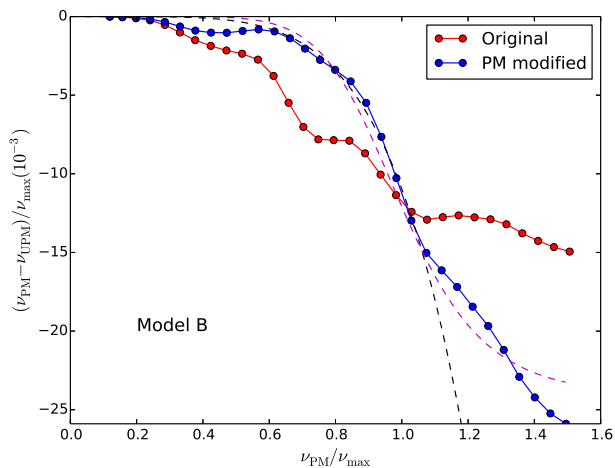}
  \caption{Frequency difference between PM and UPM as a function of the reduced frequency (the red curve with dots, same as the bottom panel of Fig. \ref{fig:dnu}). For the blue one, the value of $V_g$ is modified in PM as shown with the blue curve in the top panel of Fig. \ref{fig:Tg-Vgcs}. The dashed black and magenta lines are fitting curves with Eqs. (\ref{eq:K08}) and (\ref{eq:Lorentzian}). The 
resulting 
coefficients are $(a,b)=(-1.10\times 10^{-2}, 5.21)$ for Eq. (\ref{eq:K08}) and $(\alpha/2, \beta)=(-1.21\times 10^{-2}, 8.00)$ for Eq. (\ref{eq:Lorentzian}).}
  \label{fig:dnu_exp}
\end{figure}

To demonstrate that the elevation is responsible for the lower 
values 
of $b$ and $\beta$, we 
performed
a numerical experiment as follows: 
We took  
the mean of the quantity $V_g$ (defined as $V_g \equiv GM_r/(rc^2)$, see Fig.~\ref{fig:Tg-Vgcs} top panel) between PM and UPM and replace the original $V_g$ with it in PM. This treatment reduces the difference in the structure, 
so that 
the situation becomes similar to the models with lower effective temperature or with higher surface gravity. 
Figure
\ref{fig:dnu_exp} shows the frequency difference between UPM and the modified PM. As we can see, the difference in the frequency of the low-order modes is reduced, and the profile is similar to 
Model
A (see Fig. \ref{fig:dnu}, top panel). The fit of the parameters indicates that the value of $b$ is raised from 2.09 to 5.21
and  
that of $\beta$ from 2.56 to 8.00. 

Finally, we note that the entropy difference between the photosphere and the superadiabatic layer becomes larger with increasing effective temperature or decreasing surface gravity \citep[e.g.,][]{Ludwig99}. The entropy difference should then correlate with the superadiabaticity, 
hence with the turbulent pressure. Then, it is natural that the trend of the entropy difference shown in Fig. 4 in their paper is similar to that of the coefficients, $a$, $\alpha$, $b$, and $\beta$.

\subsection{Oscillatory features in profiles of frequency difference}
\label{sec:glitch}
As already mentioned, the Lorentzian formulation can 
reproduce the profiles of the frequency difference
properly. 
However, for some models, oscillatory features are 
conspicuously 
close to $\nu_{\rm max}$. They can lead to deviation from the Lorentzian curves. Here, we aim to interpret the oscillatory features.

A steep change in the structure or a discontinuity in stars, a so-called ``acoustic glitch'', causes modulations in spacings of frequencies \citep[e.g.,][]{Houdek07, Houdek11, Mazumdar14}. These modulations, then, may lead to modulations in the frequency difference. A useful diagnostic is the second difference with respect to the radial order $n$, defined by
\begin{eqnarray}
  \Delta_2\nu_{n,l}\equiv\nu_{n-1,l}-2\nu_{n,l}+\nu_{n+1,l},
\end{eqnarray}
where $l$ is the angular degree.
This second difference is less sensitive to the surface regions than the frequency $\nu$ itself and than the large separation $\Delta\nu=\nu_{n,l}-\nu_{n-1,l}$. Then, it is advantageous for extracting the effects of the glitches, located in the deeper interior.

\begin{figure*}
  \begin{center}
    \includegraphics[width=0.75\hsize]{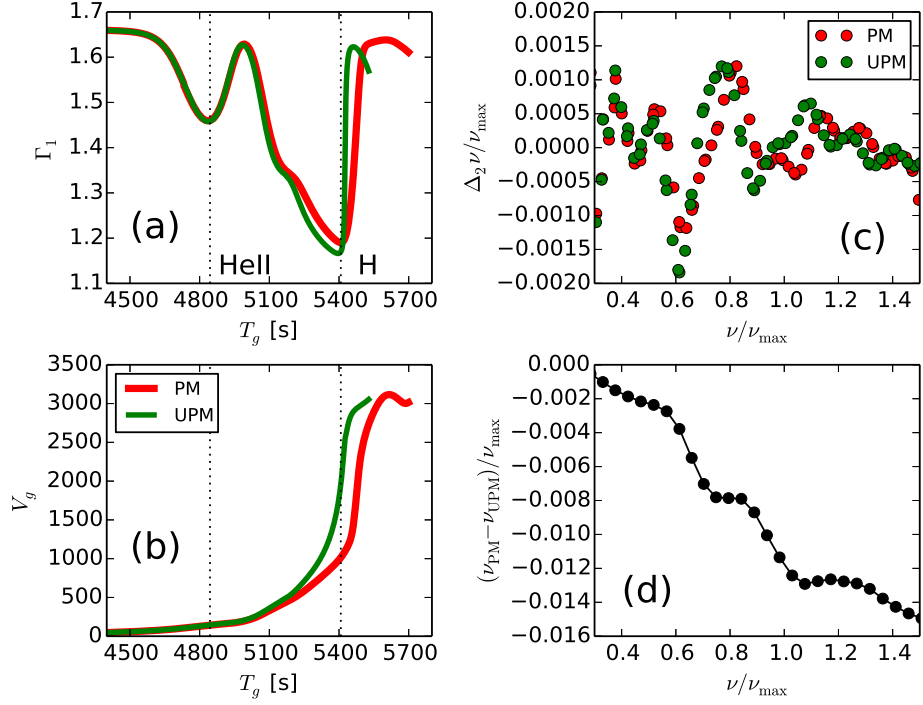}
  \end{center}
  \caption{Profiles of the original PM and UPM for 
Model
B. Panels (a) and (b) show $\Gamma_1$ and $V_g$, respectively, as functions of the acoustic radius $T_g$. The dotted vertical lines indicate the locations of the $\Gamma_1$ bumps due to the HeII and the H ionizations. Panels (c) and (d) show the second difference for $\ell=$0--2 and the frequency difference between PM and UPM for the radial modes, respectively, as functions of the frequency.}
  \label{fig:original}
\end{figure*}

\begin{figure*}
  \begin{center}
    \includegraphics[width=0.75\hsize]{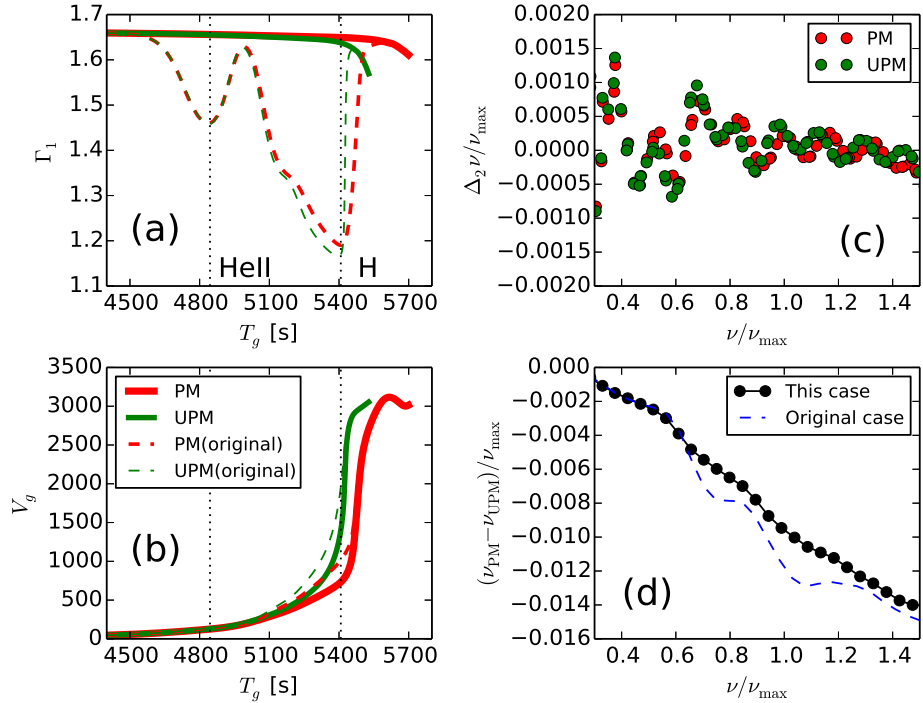}
  \end{center}
  \caption{Same as Fig. \ref{fig:original}, but for the case where the $\Gamma_1$ profiles are smoothed in both the H and the HeII bumps. In 
Panels 
(a) and (b), the profiles of the original models are shown as 
dashed lines. Similarly, the frequency difference for the original models is plotted as 
a 
dashed line in 
Panel 
(d).}
  \label{fig:smHeH}
\end{figure*}

Figure~\ref{fig:original} shows the original properties of the PM and UPM for 
Model
B. In panels (a) and (b), the $\Gamma_1$ and $V_g$ profiles of PM deviate from that of UPM above the matching point, respectively. The acoustic cavity of the PM is found 
to be
larger due to turbulent elevation. Panel (c) shows the second difference of modes with $l=$0--2. Both PM and UPM low-degree modes show modulation due to acoustic glitches. Furthermore, we can see a phase shift between them. This  
shift is responsible for 
modulating 
of the frequency difference.

Figure~\ref{fig:smHeH} shows the case where the $\Gamma_1$ profiles are smoothed 
in both 
the HeII and the H bumps [panel (a)]. Among the input variables for ADIPLS, $V_g$ and $A(\equiv \diff\ln P/\diff\ln r/\Gamma_1 - \diff\ln\rho/\diff\ln r)$ 
include 
$\Gamma_1$. In this experiment, however, we only modify $\Gamma_1$ in $V_g$, and not $A$ to avoid changing the convective stability. Then, the curves of $V_g$ form steeper walls and 
get 
closer in the upper part of the H bump [panel (b)]. This reduces the difference in the acoustic cavities between PM and UPM, 
hence the difference in the locations of the glitches. Therefore, the phase shift in $\Delta_2\nu$ becomes smaller [panel (d)].

In addition, this treatment results in the amplitude of the modulation in the second difference becoming smaller and similar between PM and UPM in the range $0.6\la\nu/\nu_{\rm max}\la 1.2$ [panel (c)]. Consequently, the oscillatory features disappear in the frequency difference.

To check which bump mainly contributes to the oscillatory features, we also 
analyzed 
the case where either the HeII or the H bump is smoothed. When we 
smoothed 
the HeII bump, we 
found 
that the oscillatory features remain unchanged. Indeed, the HeII ionization zone is located below the matching point. Then, although the modulation amplitude of $\Delta_2\nu$ is certainly reduced, the smoothing does not have any influence on the phase shift. On the other hand, when we smooth the H bump, the curve of the frequency difference becomes almost identical to the one in the panel (d) of Fig. \ref{fig:smHeH}. The locations of the glitches in the H ionization zone 
are 
different between PM and UPM 
owing 
to the turbulent elevation. Then, the smoothing reduces the phase shift, as well as the modulation amplitude. Therefore, we conclude that the oscillatory features in the frequency difference 
are 
mainly produced by the H bump and the fact that the location of the H bump is different in the PM and UPM.

\subsection{Dependence on the fitting range}
\label{sec:fitrange}

Oscillation modes are observed in a limited range, so that the power-law correction has not been used over a wide range. \cite{Kjeldsen08} 
suggest 
that the power-law index depends substantially on the frequency range considered. This means that the results of the coefficients used in the fit certainly can vary with the range.

To investigate this variation, we 
carried 
out a least-square fitting of the power law and the Lorentzian formulation 
by
changing the fitting range. We 
fixed 
the width of the frequency range to 0.8 in unit of $\nu/\nu_{\rm max}$ and 
tested 
with the ranges 0.0 to 0.8, 0.1 to 0.9, ... , and 0.8 to 1.6. 

\begin{figure}
  \begin{center}
    \includegraphics[width=\hsize]{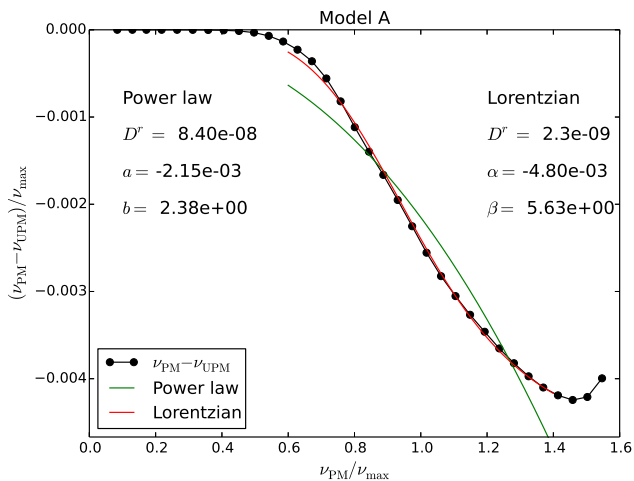}
  \end{center}
  \caption{Functional fittings with the power law (Eq. \ref{eq:K08}, the green curve) and the Lorentzian formulation (Eq. \ref{eq:Lorentzian}, the red curve) to the frequency difference of PM and UPM for 
Model 
A in the range $0.6<\nu/\nu_{\rm max}<1.4$.}
  \label{fig:dnu_0.6to1.4}
\end{figure}

\begin{figure}
  \begin{center}
    \includegraphics[width=\hsize]{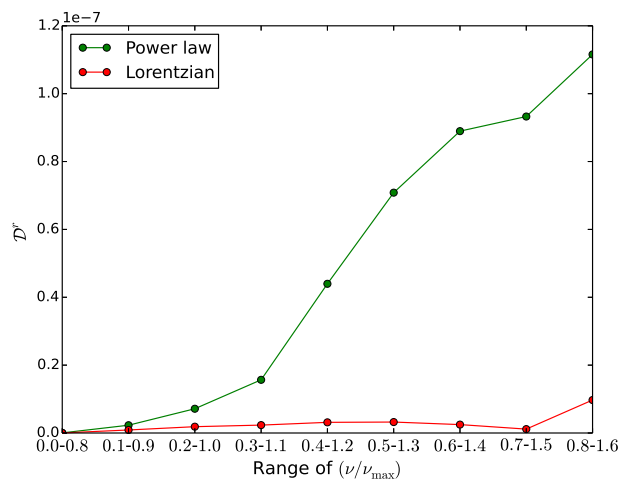}  
  \end{center}
  \caption{Dependence of the deviation 
in 
the power law (Eq. \ref{eq:K08}) and the Lorentzian formulation (Eq. \ref{eq:Lorentzian}) on the fitting range for 
Model
A.}
  \label{fig:dev_dep}
\end{figure}

Figure~\ref{fig:dnu_0.6to1.4} shows the case where the fitting is performed in the range $0.6<\nu/\nu_{\rm max}<1.4$. We see that the Lorentzian formulation is found to be more robust than the power law even if we limit the fitting range. Figure~\ref{fig:dev_dep} shows the reduced deviation ${\cal D}^r$, defined in section \ref{sec:param_fit}, of the fitting functions from the computed frequencies for 
Model
A as a function of the fitting range. For this model, the Lorentzian formulation always has
a
smaller deviation. 
This 
conclusion is valid for the other models.

\begin{figure}
  \begin{center}
    \includegraphics[width=\hsize]{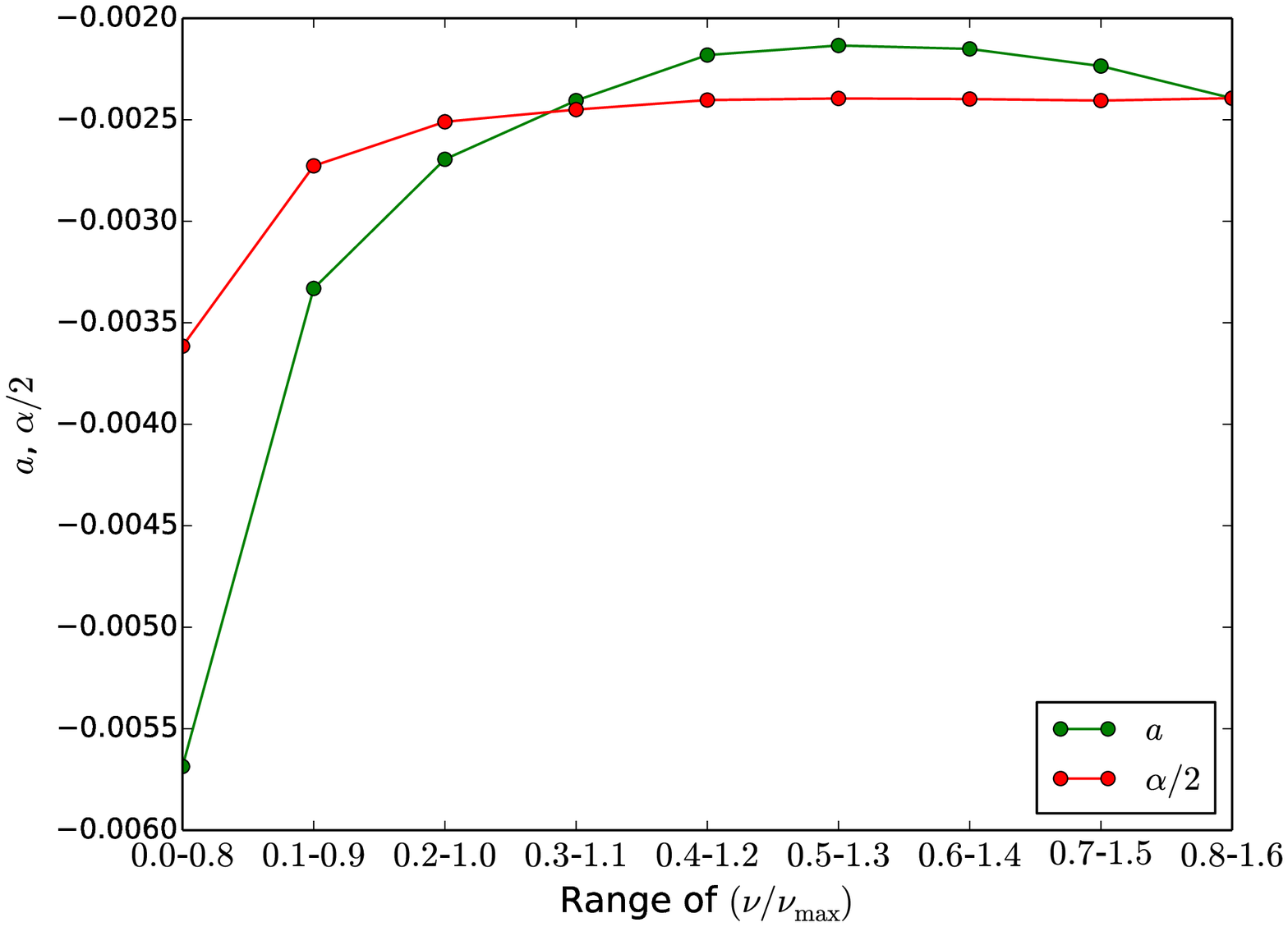}
    \includegraphics[width=\hsize]{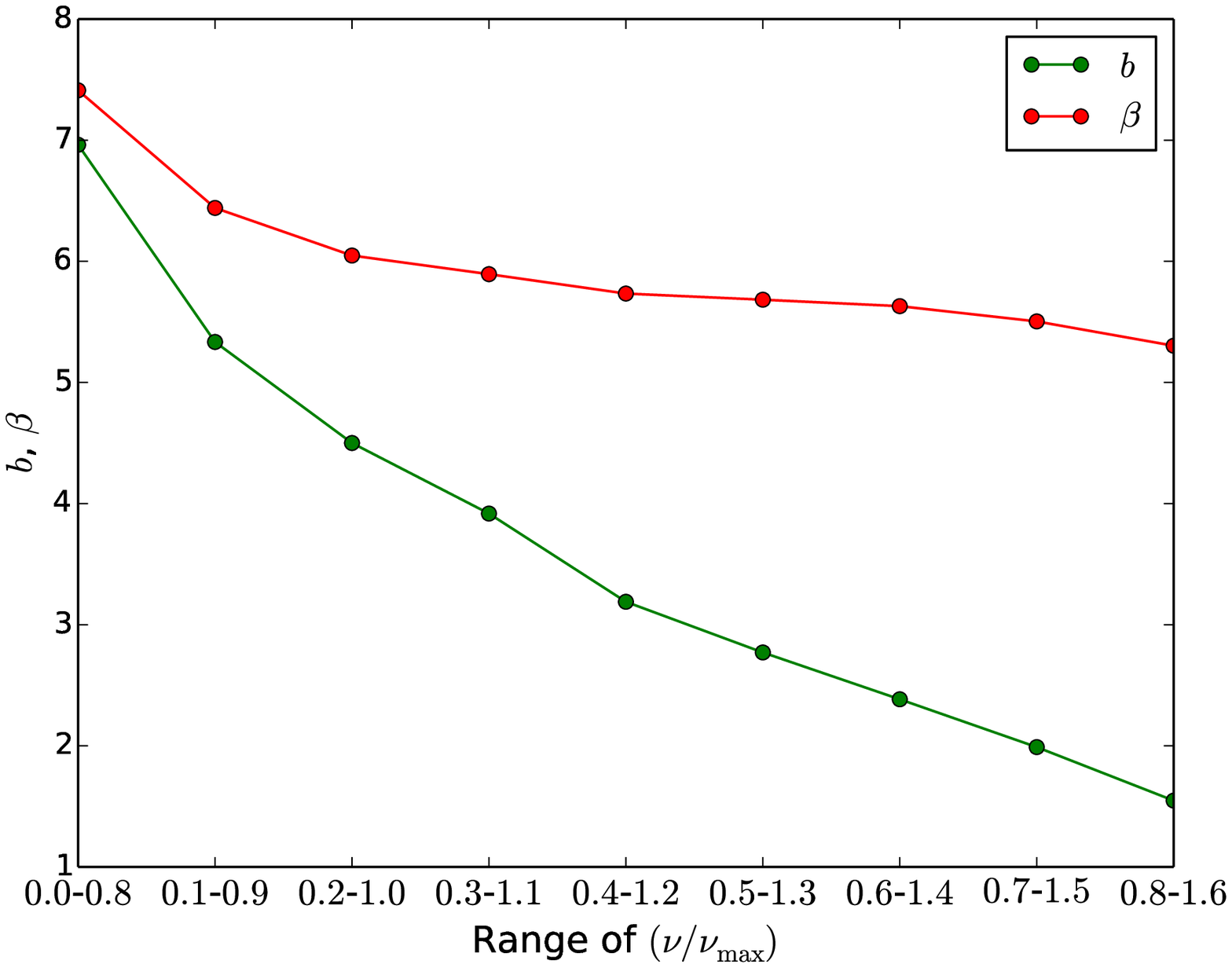}
  \end{center}
  \caption{Dependence of the coefficients of the power law (Eq. \ref{eq:K08}) and the Lorentzian formulation (Eq. \ref{eq:Lorentzian}) on the fitting range for 
Model
A.}
  \label{fig:coeff_dep}
\end{figure}

Figure~\ref{fig:coeff_dep} shows the coefficients of the power-law and the Lorentzian formulation as functions of the fitting range in the case of 
Model
A. For most models, the coefficients $a$ and $\alpha$ increase as the fitting range shifts to higher frequency, while $b$ and $\beta$ show the opposite trend. Then, the power indices $b$ and $\beta$ become higher in the lower frequency range. For 
Model
A, we find that $a$ ($\alpha$) changes by a factor of 2.67 (1.51), with the change in the frequency range from 0.0 -- 0.8 to 0.8 -- 1.6, while $b$ ($\beta$) by a factor of 4.50 (1.40). For all of the models, the coefficients of the Lorentzian formulation show 
less dependence on
the frequency range. It implies that the Lorentzian formulation is more robust than the power law for
modeling 
the surface frequency corrections. 

\section{Conclusion}\label{sec:conclusion}

We analyzed frequency differences between standard models (unpatched models, UPMs) and patched models (UPMs). The latter were constructed using the CIFIST grid \citep{Ludwig09} of 3D hydrodynamical simulation computed with the CO$^5$BOLD code \citep{Freytag12}.

We addressed the variation 
in 
the free parameters introduced by the \cite{Kjeldsen08} empirical surface-effect correction. We found that the coefficients vary significantly across the HR diagram. As a result of the functional fitting to the frequency difference, 
index $b$ decreases with increasing effective temperature or with decreasing surface gravity, while 
coefficient $a$ shows the opposite trend. These trends are caused by the elevation of the outer layers related to the modification of the hydrostatic equilibrium by the turbulent pressure. So far, $b$ has been set to the 
solar-calibrated
value in many applications, even for different stars, and $a$ has been determined by using this value of $b$. The results of this study imply that such a treatment is not appropriate. 

In addition, we confirm that the power-law 
function 
is not suitable for the high-frequency range, since the profile of the frequency difference becomes less steep as the frequency exceeds $\nu_{\rm max}$. In general, solar-like oscillations are distributed symmetrically around $\nu_{\rm max}$ in the power spectra, and we cannot neglect the modes above $\nu_{\rm max}$. Then, we propose a formulation based on a Lorentzian, which is found to successfully fit the profile of the frequency difference in the whole frequency range. The coefficients $\alpha$ and $\beta$ have similar trends as $a$ and $b$ of the power law, respectively. Moreover, we show that the Lorentzian function is more robust against both the choice of the range of corrected frequencies 
and
the glitches in the frequency differences induced by the difference in the location of the hydrogen ionization region between the PMs and UPMs.

In this paper, 
we 
limited 
ourselves to adiabatic oscillations
as a first step. 
Therefore, the surface effects are mainly caused by the turbulent pressure in the equilibrium structure. Indeed, the turbulent pressure elevates the outer layers and eventually increases the stellar radius. It leads to an expansion of the acoustic cavity, and thus to the decrease 
in 
the frequencies.

However, the effect of both nonadiabaticity and time-dependent processes between convection and oscillation were not considered. The nonadiabaticity is also expected to play a non-negligible role in the near-surface regions, since the thermal timescale is comparable to or shorter than the oscillation periods. Moreover, convection can also affect oscillations through time-dependent processes. For instance, the perturbation of turbulent pressure and of the convective heat flux must be related to the nonadiabatic processes of the oscillations. 
Computing
nonadiabatic oscillations using a time-dependent treatment of convection is thus a natural way to investigate these mechanisms. We will pursue 
these subjects 
in a subsequent paper. 

\begin{acknowledgements}
  The authors are grateful to M.-J. Goupil and M.-A. Dupret for their helpful comments. H.G.L. acknowledges financial support by the Sonderforschungsbereich SFB 881 ``The Milky Way System'' (subproject A4) of the German Research Foundation (DFG). T.S. has been supported by the ANR program IDEE ``Interaction Des \'Etoiles et des Exoplan\`etes'' (Agence Nationale de la Recherche, France).
\end{acknowledgements}

%-------------------------------------------------------------------
\bibliographystyle{aa}
\bibliography{sonoi}

\end{document}